\documentclass[twocolumn,showpacs]{revtex4-1}
\usepackage[utf8x]{inputenc}

\begin{document}
\title{Growth of covariant perturbations in the contracting phase of a bouncing universe}
\author{Atanu Kumar \footnote{atanu.kumar@saha.ac.in}}
\affiliation{Theory Division, Saha Institute of Nuclear physics\\
1/AF, Bidhannagar, Kolkata 700098, India\\
}

\newcommand{\fa}{\frac{1}{2}}
\newcommand{\fb}{\frac{1}{3}}
\newcommand{\fc}{\frac{2}{3}}
\newcommand{\du}{\dot{u}}
\newcommand{\mc}{\mbox{Curl}}
\newcommand{\lp}{\tilde{\nabla}^2}

\begin{abstract}
In this paper we examine the validity of the linear perturbation theory near a bounce in the covariant analysis. Some linearity parameters are defined to set up conditions for a linear theory. Linear evolution of density perturbation and gravitational waves have been computed previously. We have calculated the vector and scalar induced parts of the shear tensor. For radiationlike and dustlike single fluid dominated collapsing Friedmann-Lema\^{\i}tre-Robertson-Walker background it is shown that the linearity conditions are not satisfied near a bounce.
\end{abstract}

\maketitle


\section{Introduction}

The cosmological models for a collapsing universe approaching an expanding one through a nonsingular bounce or its cyclic repetitions are being studied over a long time \cite{Tolman}-\cite{Biswas:2011ar}. For example, in a dustlike scalar field dominated contracting Friedmann-Lema\^{\i}tre-Robertson-Walker (FLRW) background both the curvature perturbation $\zeta$ and the Bardeen potential $\Phi$  have been observed to grow, whereas in an expanding phase $\zeta$ remains constant and $\Phi$ decays \cite{Finelli:2001sr}. Such growing modes of the perturbations raise doubts on the validity of the linear perturbation theory close to the bounce \cite{Lyth:2001nv}. However, some recent papers \cite{Allen:2004vz}, \cite{Vitenti:2011yc} claim that there exists a choice of gauge in which the linear approximations remains valid, whereas in other commonly defined gauges such approximations become invalid.

The power spectra of $\Phi$ and $\zeta$ are given in \cite{Vitenti:2011yc} for a contracting phase (just before the bounce) and for an expanding phase (long after the bounce). $\Phi$ has a constant mode and a growing mode close to the bounce. For example, in a dust dominated contracting flat FLRW background, $\Phi$ behaves as
\begin{equation}
 \Phi_k(\eta) = C_1(k)+C_2(k)\eta^{-5}
\end{equation}
where $\eta$ is the conformal time and $a$ is the scale factor, $a\propto (-\eta)^2$. Due to the growing mode, $\Phi$ becomes very large near the bounce. On the other hand, although $\zeta$ grows near the bounce, its amplitude remains always less than the constant mode of $\Phi$. If the perturbations are assumed to be small at some initial time far away from the bounce then the constant mode of $\Phi$ and hence $\zeta$ remain small near the bounce. In order to examine the effect of these growing modes of $\Phi$ and $\zeta$ on the validity of linear approximations, one needs to compare metric perturbations with their background values. Metric perturbation about an FLRW background is written as
\begin{equation}
 g_{\mu\nu}=g^{0}_{\mu\nu}+h_{\mu\nu}
\end{equation}
Let us consider the scalar perturbation $\phi$, defined by $g_{00}=-1+2\phi$. A necessary condition for the perturbation to be linear is that $\phi\ll 1$. In the Newtonian gauge, $\phi$ is equal to the Bardeen potential $\Phi$. Now from the behavior of $\Phi$ it is evident that $\phi \gg 1$, close to the bounce. On the other hand, in the uniform curvature gauge $\phi=\frac{3}{2}(1+w)\zeta$ in a single fluid dominated phase. Since $\zeta$ is small close to the bounce, $\phi$ also remains small. It shows that the perturbation remains linear close to the bounce if one uses the uniform curvature gauge instead of the Newtonian gauge. Similarly, {\em at the bounce}, where Hubble's constant goes to zero, neither the Newtonian nor the uniform curvature gauge but the synchronous gauge preserves the linearity of the perturbation. This shows that there exists at least one gauge in which the perturbations remain small near bounce so that linear perturbation theory holds. But in other gauges perturbations grow.

The purpose of this paper is to analyze the problem in a fully covariant approach first proposed by Hawking \cite{Hawking} and further developed in \cite{Ellis:1989jt}. In this approach, the dynamical variables are fully gauge invariant and hence, the analysis is completely independent of the choice of gauge conditions. Moreover, in contrast with the perturbative expansion of Einstein equations used in standard perturbation theory, the dynamical equations are exact and nonlinear. One can relate the variables used in the covariant analysis with the gauge invariant variables used in the standard perturbation theory. For example, the vector $\zeta_a$, defined in Eq.~(\ref{zetaa}) of Sec.~\ref{variables}, can be seen as a generalization of the curvature perturbation $\zeta$, used in linear theory. In particular $\zeta_a$ coincides with the usual $\zeta$ for long wavelengths, but the two quantities differ on small scales, where spatial derivatives cannot be neglected. 

We consider collapsing radiation and dust dominated FLRW backgrounds. First, the linear perturbation equations are solved in these backgrounds near the bounce. Then these solutions are used to compare the linear and nonlinear terms in the full nonlinear perturbation equations to investigate the validity of the linearized approximations. Our discussion is general and not specific to a particular model of bounce.

So far it has been checked whether perturbations are sufficiently small compared with their background quantities. If that criterion holds, second order perturbations are assumed to be even smaller and linear perturbation theory is considered to be a good approximation. But this test will not work in covariant formalism as the background values of the gauge invariant perturbations are zero. In the present work, we have checked whether higher order perturbations are truly smaller compared with 1st order ones. We will show that the answer is not positive for all perturbation modes near bounce.

The outline of this paper is as follows. In Sec.~\ref{covariant}, we set up our notations and give a brief introduction to the covariant perturbation theory. In Sec.~\ref{variables} we list all the gauge invariant variables. In Sec.~\ref{equations} we give the full nonlinear perturbation equations and their linearized versions close to a flat FLRW background. Also the conditions for validity of the linear approximation are set up. In Sec.~\ref{solution} we present the solutions of the linearized equations for a radiation and dust dominated background and finally, in Sec.~\ref{compare} we compare the first and higher order terms in full nonlinear equations.


\section{Theory of covariant cosmological perturbation}
\label{covariant}

In the standard perturbation theory \cite{Bardeen:1980kt},\cite{Mukhanov:1990me}, \cite{Weinberg:2008zzc} we consider an idealized model of universe $\bar{S}$, usually a FLRW  one with metric $\bar{g}_{ab}$, energy density $\bar{\mu}$, pressure density $\bar{p}$, etc. Then we perturb the model to obtain a realistic universe $S$ with corresponding quantities $g_{ab}$, $\mu$ and $p$. The perturbations at each space-time point $q$ is defined as the difference of these quantities at $q$.
\begin{equation}
 \delta g_{ab} = g_{ab}-\bar{g}_{ab},\quad \delta\mu=\mu-\bar{\mu}, \quad \delta p=p-\bar{p}.
\end{equation}
But $S$ and $\bar{S}$ have different manifold structure. Mathematically these means we have to define a map $\Phi$ from $\bar{S}$ to $S$. Then we can pull back each quantity $\mu$ in $S$ to its image $\Phi^*\mu$ in $\bar{S}$.
\begin{equation}
 \Phi^*\mu(\bar{q})=\mu(\Phi(\bar{q})), \quad \bar{q}~\epsilon~\bar{S}, \quad q=\Phi(\bar{q})~\epsilon ~S.
\end{equation}
Then the perturbation is defined as
\begin{equation}
 \delta\mu(\bar{q})=\Phi^*\mu(\bar{q})-\bar{\mu}(\bar{q}).
\end{equation}
There are coordinate freedoms in both the manifolds $S$ and $\bar{S}$ and these choices determine the map $\Phi$. These arbitrariness in the $\Phi$ makes the definition of $\Phi^*\mu(\bar{q})$ and hence $\delta\mu$ ambiguous. This is the gauge freedom in $\mu$. To study the evolution of perturbations we have to fix the gauge i.e. specify the map $\Phi$, then there will be no ambiguity in $\delta\mu$.

But, $S$ is the realistic universe and $\bar{S}$ is our theoretical idealization. Given a physical universe $\bar{S}$ we can not construct the background model $S$ uniquely without any further information. 

In \cite{Hawking}, \cite{Ellis:1989jt} an alternative representation to standard perturbation theory is given. This representation involves quantities, whose background values are zero. Suppose $f$ is such a quantity, defined on $S$. Then we can consider $f$ itself as the perturbation variable and it is completely gauge invariant. Consequently, the evolution equations for $f$ are fully covariant. The quantities, which do not vanish on background space-time, are called zeroth order variables. By the statement, ``The background space-time is flat FLRW," we mean that zeroth order quantities are governed by the Friedmann equations with zero 3-curvature.

Now we consider a general perfect fluid flow in curved space-time \cite{Ellis:1973},\cite{Ellis:1998ct}. In cosmological context, there will always be a preferred family of world lines representing the motion of typical observers. Let the four velocity of those fundamental observers be $u^a$.
\begin{equation}
 u^a=\frac{dx^a}{d\tau}, \quad u^au_a=-1,
\end{equation}
where $\tau$ is the proper time along the fundamental world lines. The metric of the 3-space, orthogonal to $u^a$, is
\begin{equation}
 h_{ab}=g_{ab}+u_au_b~\Rightarrow h^a_{~b}h^b_{~c}=h^a_{~c}~,~~h_{ab}u^b=0.
\end{equation}
Time derivative of any tensor $S^{a...}_{~b...}$ along fluid flow lines is defined as
\begin{equation}
 \dot{S}^{a...}_{~b...} = u^c\nabla_c S^{a...}_{~b...}.
\end{equation}
The first covariant derivative of four velocity can be written as
\begin{equation}
 \nabla_b u_a = \frac{1}{3}\theta h_{ab}+\sigma_{ab}+\omega_{ab}-u_b \du_a,
\end{equation}
Where the trace $\nabla^au_a=\theta$ is the expansion, traceless symmetric part, $\sigma_{ab}=\sigma_{(ab)}$ is the shear tensor, antisymmetric part, $\omega_{ab}=\omega_{[ab]}$ is the vorticity and $\dot{u_a}$ is the acceleration. We can define a scale $a(\tau)$ along each world line as
\begin{equation}
 \label{theta} \theta = 3\frac{\dot{a}}{a},
\end{equation}
$a$ can be determined up to a multiplicative constant.
\begin{equation}
 \sigma_{ab}u^b=0=\omega_{ab}u^b
\end{equation}
The space-time curvature (Riemann) tensor $R_{abcd}$, defined by $2\nabla_{[a}\nabla_{b]}v_c=R_{abcd}v^d$ is made up of Ricci tensor $R_{ab}$ and the tracefree Weyl tensor components $C_{abcd}$. 
\begin{equation}
 R_{abcd}=C_{abcd}+g_{a[c}R_{d]b}-g_{b[c}R_{d]a}-\frac{1}{3}g_{a[c}g_{d]b}R
\end{equation}
The trace part $R_{ab}$ is determined by the Einstein equation
\begin{equation}
 R_{ab}-\frac{1}{2}Rg_{ab}=\kappa T_{ab}.
\end{equation}
$\kappa=8\pi G$ is the Plank length squared. The energy momentum tensor $T_{ab}$ for a perfect fluid can be written in terms of energy density $\mu$, pressure density $p$ and $u^a$,
\begin{equation}
 T_{ab}=(\mu+p)u_au_b+pg_{ab}=\mu u_au_b + ph_{ab}.
\end{equation}
The energy-momentum tensor must satisfy the conservation equation,
\begin{equation}
\label{consv} \nabla_b T^{ab}=0.
\end{equation}


\section{Gauge invariant variables}
\label{variables}
 A FLRW model is a perfect fluid space-time characterized by the conditions \cite{Ellis:1989jt},
\begin{equation}
 \sigma_{ab}=\omega_{ab}=\dot{u^a}=0,
\end{equation}
which implies
\begin{equation}
 \mu=\mu(t),\quad p=p(t),\quad \theta=\theta(t),
\end{equation}
where $t$ is the cosmic time, defined by $u_a=-\nabla_a t$; and, Weyl tensor vanishes
\begin{equation}
 C_{abcd}=0.
\end{equation}
So this space-time is conformally flat.

From the above characterization, we can list some simple gauge invariant quantities \cite{Bruni:1992dg}:\\
(1) Shear, vorticity and acceleration,
 \begin{eqnarray}
\label{shear}  \sigma_{ab} &=& (h_{(a}^{~c}h_{b)}^{~d}-\frac{1}{3}h_{ab}h^{cd})\nabla_d u_c, \\
  \omega_{ab} &=& h_{[a}^{~c}h_{b]}^{~d}\nabla_d u_c, \\
  \dot{u}_a &=& u^b\nabla_b u_a.
 \end{eqnarray}
Rotation vector is defined as $\omega_a=\frac{1}{2}\epsilon_{abc}\omega^{bc}$. $\epsilon_{abc}$ is Levi-Civita tensor in 3-hypersurface
defined by $\epsilon_{abc}=\eta_{abcd}u^d$. \\
(2) ``Electric'' and ``magnetic'' parts of the Weyl tensor,
   \begin{eqnarray}
    E_{ab}=C_{acbd}u^cu^d, \quad  H_{ab}=\frac{1}{2}C_{acpq}\eta^{pq}_{~~bd}u^cu^d.
   \end{eqnarray}
(3) Spatial gradients of the energy density, pressure density and expansion,
 \begin{eqnarray}
\label{XYZ}  X_a=\kappa h_a^{~b}\nabla_b\mu ,\quad Y_a=\kappa h_a^{~b}\nabla_b p ,\quad Z_a=h_a^{~b}\nabla_b\theta.
 \end{eqnarray}
There are many other gauge invariant quantities, which vanish in the background model. We should mention two of them those will appear in the evolution equations.
\begin{equation}
\label{AA} A=\nabla^a \dot{u}_a, \quad A_a=h_{a}^{~b}\nabla_bA.
\end{equation}

It is evident that $\mu$, $p$ and $\theta$ are zeroth order variables, whereas those defined in (\ref{shear})-(\ref{AA}) are called first order variables. We can construct higher order variables from first order ones as
\begin{eqnarray}
 \omega^2= \frac{1}{2}\omega_{ab}\omega^{ab},~~ \sigma^2=\frac{1}{2}\sigma_{ab}\sigma^{ab},~~ \sigma_{ab}E^{ab},~~ \sigma_{ab}H^{ab}~~.
\end{eqnarray}

The comoving fractional density gradient, defined as $\mathcal{D}_a=a\frac{X_a}{\kappa\mu}$,  is similar to the variable $\delta=\delta\mu/\mu$, used in standard perturbation theory and is observable in principle. So we will concentrate on the equations of $X_a$. The gauge invariant quantities defined in (\ref{shear})-(\ref{XYZ}) form a closed set of equations, as will be seen in the next section.

To demonstrate the relation between covariant and coordinate based approaches let us define another variable $\zeta_a$ as \cite{Langlois:2005ii}, \cite{Langlois:2010vx}
\begin{equation}
\label{zetaa} \zeta_a = W_a + \frac{X_a}{3\kappa(\mu+p)}, \quad W_a =h_a^{~b}\nabla_b \log a.
\end{equation}
A cosmological space-time, close to the flat FLRW geometry can be described by the metric in comoving coordinates,
\begin{eqnarray}
 & & ds^2 = -(1+2\phi)dt^2 + 2a(\partial_i B +S_i)dtdx^i ~~\nonumber \\
& &~ +a^2((1+2\psi)\delta_{ij}+2\partial_i\partial_jE+2\partial_{(i}F_{j)}+h_{ij})dx^idx^j.~~
\end{eqnarray}
The matter content is a perfect fluid with energy density $\mu(t,x^i)=\bar{\mu}(t)+\delta\mu(t,x^i)$, pressure $p(t,x^i)=\bar{p}(t)+\delta p(t,x^i)$ and 4-velocity $u^{\mu}=\bar{u}^{\mu}+\delta u^{\mu}$.
\begin{equation}
 \delta u_0 = -\phi, \quad \delta u_i=\partial_i \delta u + \delta u^V_i.
\end{equation}
At linear order, spatial components of $\zeta_a$ are
\begin{equation}
 \zeta_i = \partial_i \zeta^S, \quad \zeta^S = \frac{\delta a}{a} - \frac{H}{\dot{\bar{\mu}}}\delta\mu.
\end{equation}
Up to linear order $\delta a/a$ can be written as
\begin{equation}
 \frac{\delta a}{a} = \psi+\fb\nabla^2 E +\fb\int dt \nabla^2 \delta u.
\end{equation}
So, $\zeta^S$ coincides with familiar definition of comoving curvature perturbation  $\zeta = \psi - \frac{H}{\dot{\bar{\mu}}}\delta\mu$ when the spatial gradients are negligible. For adiabatic perturbations, $\zeta^S$ is conserved at all scales whereas $\zeta$ is conserved only for large wavelength modes.


\section{Dynamic equations and constraints}
\label{equations}


\subsection{Exact equations}

The time and space components of the energy-momentum conservation equation (\ref{consv}) leads to
\begin{eqnarray}
\label{energy} \dot{\mu}+\theta(\mu+p) &=& 0, \\
\label{momentum} \kappa(\mu+p)\dot{u}_a+Y_a &=& 0.
\end{eqnarray}
The Raychaudhury equation gives the evolution of $\theta$ along fluid flow lines \cite{Ehlers:2006aa},\cite{Raychaudhuri:1953yv}
\begin{equation}
\label{ray} \dot{\theta}+\fb\theta^2-A+\frac{1}{2}\kappa(\mu+3p)+2(\sigma^2-\omega^2) =0.
\end{equation}
To solve the zero order background equations (\ref{energy}) and (\ref{ray}) we need the matter equation of state in the form $p=p(\mu)$. Then (\ref{XYZ}) and (\ref{momentum}) imply that $\dot{u}_a$ and $Y_a$ can be written in terms of $X_a$.

The equations for $X_a$ and $Z_a$ are given in \cite{Ellis:1989jt}
\begin{eqnarray} 
\label{xdot} a^{-4}h_{a}^{~b}(a^4X_b\dot{)}&=& -\kappa(\mu+p)Z_a - (\omega^b_{~a}+\sigma^b_{~a})X_b, \\
 a^{-3}h_{a}^{~b}(a^3Z_b\dot{)}&=& \mathcal{R}\dot{u}_a - \fa X_a + A_a + 2h_{a}^{~b}\nabla_b(\omega^2-\sigma^2) \nonumber \\
\label{zdot} & & \;\;\;\;\;\;\;\;\;\;\;\;\;\;\;\;\;\;\;\;\;\;\; - (\omega^b_{~a}+\sigma^b_{~a})Z_b,
\end{eqnarray}
where, 
\begin{equation}
 \label{r} \mathcal{R} = \kappa\mu -\fb\theta^2+A+2(\omega^2-\sigma^2).
\end{equation}
Equations for four other variables $\omega_{ab}$, $\sigma_{ab}$, $E_{ab}$ and $H_{ab}$ are given in \cite{Hawking}
\begin{widetext}
\begin{eqnarray}
\label{odot} a^{-2}h_{a}^{~c}h_{b}^{~d}(a^2\omega_{cd}\dot{)} &=& h_{a}^{~c}h_{b}^{~d}\nabla_{[d}\du_{c]} + 2\sigma_{c[a}\omega_{b]}^{~c}, \\
\label{sdot} a^{-2}h_{a}^{~c}h_{b}^{~d}(a^2\sigma_{cd}\dot{)} &=& -E_{ab} + \nabla_{\langle b}\du_{a\rangle} - \omega_{ac}\omega^c_{~b} -\sigma_{ac}\sigma^c_{~b}+\frac{2}{3}h_{ab}(\sigma^2-\omega^2)+\du_a\du_b, \\
\label{edot} a^{-3}h_{a}^{~c}h_{b}^{~d}(a^3E_{cd}\dot{)} &=& - \mc H_{ab} -\fa\kappa(\mu+p)\sigma_{ab} + E^c_{~(a}\omega_{b)c} + E^c_{~(a}\sigma_{b)c} + \epsilon_{acd}\epsilon_{bpq}\sigma^{cp}E^{dq} - 2H^c_{~(a}\epsilon_{b)cd}\du^d, \\
\label{hdot} a^{-3}h_{a}^{~c}h_{b}^{~d}(a^3H_{cd}\dot{)} &=& \mc E_{ab} + H^c_{~(a}\omega_{b)c} + H^c_{~(a}\sigma_{b)c}+\epsilon_{acd}\epsilon_{bpq}\sigma^{cp}H^{dq}  - 2H^c_{~(a}\epsilon_{b)cd}\du^d .
\end{eqnarray}
\end{widetext}

We have used following notations \cite{Maartens:1997fg}: 
\begin{eqnarray*}
 & &\lambda_{(ab)} = \fa(\lambda_{ab}+\lambda_{ba}), \quad \lambda_{[ab]}=\fa(\lambda_{ab}-\lambda_{ba}), \\
 & & \lambda_{\langle ab\rangle} = h_{a}^{~c}h_{b}^{~d}(\lambda_{(cd)}-\fb h_{cd}\lambda^e_{~e}), \quad \\
 & & \mc \lambda_{ab} = h^e_{~(a}\epsilon_{b)cd}\nabla^d\lambda_e^{~c}.
\end{eqnarray*}

There are also constraint relations which must be satisfied at some initial time on each world line \cite{Hawking},
\begin{eqnarray}
& & h_{a}^{~c}\nabla_b(\omega^b_{~c}+\sigma^b_{~c})-\du^b(\omega_{ab}+\sigma_{ab}) = \frac{2}{3}Z_a, \\
& & \nabla_a\omega^a = 2\du_a\omega^a, \\
& & \mc\omega_{ab}+\mc\sigma_{ab} = -H_{ab}, \\
& & h_{a}^{~c}\nabla^bE_{bc}+3H_{ab}\omega^b-\epsilon_{abc}\sigma^b_{~d}H^{cd} = \fb X_a, \\
& & h_{a}^{~c}\nabla^bH_{bc}-3E_{ab}\omega^b-\epsilon_{abc}\sigma^b_{~d}H^{cd} = \kappa(\mu+p) \omega_a.
\end{eqnarray}


\subsection{Linearization about FLRW model}
\label{lineareqn}
Let us consider the universe to be almost FLRW and linearize the perturbation equation. Since $\mu$, $p$, $\theta$ are zeroth order variables, we consider only zeroth order terms of (\ref{energy}) and (\ref{ray}). Then the Raychaudhury equation becomes
 \begin{equation}
\label{raylin} \dot{\theta}+\fb\theta^2+\frac{1}{2}\kappa(\mu+3p)=0.
\end{equation}
Now $a$ can be interpreted as the Robertson-Walker scale factor. Linearized forms of the perturbation equations become
\begin{eqnarray}
\label{lxdot}  a^{-4}(a^4X_a\dot{)} &=& -\kappa(\mu+p)Z_a, \\
\label{lzdot}  a^{-3}(a^3Z_a\dot{)} &=& - \fa X_a + A_a, \\
\label{lodot}  a^{-2}(a^2\omega_{ab}\dot{)} &=& \nabla_{[b}\du_{a]}, \\
\label{lsdot}  a^{-2}(a^2\sigma_{ab}\dot{)} &=& -E_{ab} + \nabla_{\langle b}\du_{a\rangle}, \\
\label{ledot}  a^{-3}(a^3E_{ab}\dot{)}  &=& -\mc H_{ab} -\fa\kappa(\mu+p)\sigma_{ab}, \\
\label{lhdot}  a^{-3}(a^3H_{ab}\dot{)} &=& \mc E_{ab}. 
\end{eqnarray}
Since background FLRW universe is flat, $\mathcal{R}$ is a first order quantity. That is why the 1st term in the right hand side of (\ref{zdot}) is neglected in (\ref{lzdot}). Also we have the constraints,
\begin{eqnarray}
& & \nabla_b(\omega^b_{~a}+\sigma^b_{~a}) = \frac{2}{3}Z_a , \qquad
 \nabla_a\omega^a=0, \\
& & H_{ab} = -\mc\omega_{ab}-\mc\sigma_{ab}, \\
& & \nabla^bE_{ab} = \fb X_a, \qquad
 \nabla^bH_{ab}=\kappa(\mu+p) \omega_a.
\end{eqnarray}


\subsection{Condition for linearity}
\label{condlin}

In the standard perturbation theory, linearization is justified if the perturbations are small with respect to corresponding background quantities. The gauge invariant variables defined in Sec. \ref{variables} are nonlinear and so far we have not assumed any ``smallness." However when we consider the linear evolution, some conditions have to be imposed. They cannot be compared with corresponding background quantities, because those are zero from the definition of gauge invariant variables. The definition of dimensionless variable \cite{Goheer:2004gk} is not unique, because one can multiply any power of scale factor $a$, and that variable remains dimensionless. The natural way is to demand that higher order terms of perturbation equations remain small with respect to first order terms. 

Let us concentrate on the evolution of density perturbations $X_a$. It is observed that $X_a$ is coupled with $Z_a$ up to linear order in Eqs.~(\ref{xdot}) and (\ref{zdot}). To compare nonlinear terms of those equations with linear terms we define following parameters:
\begin{eqnarray}
 \varepsilon_1 &=& \frac{\left|\omega^b_{~a}X_b\right|}{\left|\kappa(\mu+p)Z_a\right|},\quad
 \varepsilon_2 = \frac{\left|\sigma^b_{~a}X_b\right|}{\left|\kappa(\mu+p)Z_a\right|},~ \nonumber \\
 \varepsilon_3 &=& \frac{\left|\mathcal{R}\du_a\right|}{\left|\fa X_a\right|},\quad
 \varepsilon_4 = \frac{\left|2h_a^{~b}\nabla_b\omega^2\right|}{\left|\fa X_a\right|},\quad
 \varepsilon_5 = \frac{\left|2h_a^{~b}\nabla_b\sigma^2\right|}{\left|\fa X_a\right|},~ \nonumber \\
 \varepsilon_6 &=& \frac{\left|\omega^b_{~a}Z_b\right|}{\left|\fa X_a\right|}, \quad
 \varepsilon_7 = \frac{\left|\sigma^b_{~a}Z_b\right|}{\left|\fa X_a\right|}, \nonumber \\
 \tilde{\varepsilon}_3 &=& \frac{\left|\mathcal{R}\du_a\right|}{\left|A_a\right|},\quad
 \tilde{\varepsilon}_4 = \frac{\left|2h_a^{~b}\nabla_b\omega^2\right|}{\left|A_a\right|},\quad
 \tilde{\varepsilon}_5 = \frac{\left|2h_a^{~b}\nabla_b\sigma^2\right|}{\left|A_a\right|}, \nonumber \\
\label{eps} \tilde{\varepsilon}_6 &=& \frac{\left|\omega^b_{~a}Z_b\right|}{\left|A_a\right|}, \quad 
 \tilde{\varepsilon}_7 = \frac{\left|\sigma^b_{~a}Z_b\right|}{\left|A_a\right|}.
\end{eqnarray}

The linear perturbation theory is valid for the solutions of (\ref{xdot}) and (\ref{zdot}), if the following conditions are satisfied throughout the regime under consideration:
\begin{eqnarray}
 & & \mathbf{(1)} \; \varepsilon_1, \varepsilon_2 \ll 1, \nonumber \\
 & & \mathbf{(2)} \; \varepsilon_3 , \varepsilon_4, \varepsilon_5, \varepsilon_6, \varepsilon_7 \ll 1, ~\mbox{and}/\mbox{or}~ \tilde{\varepsilon}_3, \tilde{\varepsilon}_4, \tilde{\varepsilon}_5, \tilde{\varepsilon}_6, \tilde{\varepsilon}_7 \ll 1. \nonumber
\end{eqnarray}


\section{Solution of linearized equations for specific matter description}
\label{solution}

Now we will concentrate on adiabatic perturbations in a collapsing FLRW background, for ``Dust'' and ``Radiation'' dominated cases. The solutions of  coupled equations (\ref{lxdot}) and (\ref{lzdot}) for $X_a$ and $Z_a$ are obtained independently. Then $\du_a$ is found from (\ref{momentum}). The $\omega_{ab}$ also obtained from (\ref{lodot}). Then we can solve (\ref{lsdot})-(\ref{lhdot}) using $\omega_{ab}$ and $\du_a$ as a source.

Let us consider equation of state of the dominating single fluid near bounce is
\begin{equation}
\label{eos} p=w\mu ~\Rightarrow~ Y_a=wX_a.
\end{equation}
Zeroth order quantities, obtained from (\ref{energy}), (\ref{raylin}) and (\ref{theta}) are
\begin{equation}
\label{zero} \mu=\frac{M}{a^{3(1+w)}},\quad a=\large{(}-\frac{t}{t_*}\large{)}^{\frac{2}{3(1+w)}},\quad \theta=\frac{2}{1+w}\frac{1}{t}.
\end{equation}

We assume the bounce occurs at $t=-t_b$ such that $t_b$ is of the order of Plank scale. The initial time $t=-t_*$ is well inside the regime dominated by the single fluid considered here such that, $a_b=a(-t_b)<<1$. The constants $M$ and $t_*$ have the following relation:
\begin{equation}
 \kappa M t_*^2=\frac{4}{3(1+w)^2}.
\end{equation}

The acceleration is expressed in terms of $X_a$ as
\begin{equation}
\label{uadot} \du_a = -\frac{w}{1+w}\frac{X_a}{\kappa\mu}.
\end{equation}
Now substituting (\ref{uadot}) in (\ref{lzdot}) and using the background quantities from (\ref{zero}), we extract the second order differential equation of $X_a$ from (\ref{lxdot}) and (\ref{lzdot}).
\begin{equation}
\label{ddtX} \ddot{X}_a + \frac{10+3w}{3}\theta\dot{X}_a + \frac{11+3w}{6}\theta^2X_a-wh_a^{~b}\nabla_b\nabla^c X_c =0.
\end{equation}
In principle, we can solve (\ref{ddtX}) and obtain $Z_a$ from (\ref{lxdot}). Solutions for density perturbation $\mathcal{D}_a=a\frac{X_a}{\kappa\mu}$ for radiation and dust dominated FLRW background are given in \cite{Ellis:1989ju} and \cite{Ellis:1989jt} respectively.

The $\omega_{ab}$ is easily obtained from  (\ref{lodot}). Since $X_a$ and hence $\du_a$ is spatial gradient of a scalar, the right-hand side of (\ref{lodot}) can be simplified for a torsion-free manifold.
\begin{eqnarray}
 h_a^{~c}h_b^{~d}\nabla_{[d}\du_{c]} &=& -\frac{w}{1+w}\frac{1}{\kappa\mu}h_a^{~c}h_b^{~d}\nabla_{[d}X_{c]} \nonumber \\
 &=& -\frac{w}{1+w}\frac{1}{\kappa\mu}\kappa h_a^{~c}h_b^{~d}\nabla_{[d}h_{c]}^{~e}\nabla_e\mu  \nonumber \\
 &=& -\frac{w}{1+w}\frac{\dot{\mu}}{\mu}h_a^{~c}h_b^{~d}\nabla_{[d}u_{c]} \nonumber \\
 &=& w\theta\omega_{ab}. \nonumber 
\end{eqnarray}
Substitution of this in (\ref{lodot}) results in simple evolution of $\omega_{ab}$.
\begin{equation}
\label{omega} \omega_{ab}=\Omega_{ab}a^{-2+3w},
\end{equation}
where $\dot{\Omega}_{ab}=0$. Let us consider another important quantity, 
 \begin{equation}
\label{ra}  r_a=\nabla^b\omega_{ab}.
 \end{equation}
For any first order (0,2) tensor $\lambda_{ab}$ in FLRW background, the time derivative and spatial divergence do not commute; rather, they satisfy the following relation:
\begin{equation}
 \label{comm} (h_a^{~b}\nabla^c\lambda_{bc}\dot{)}=h_a^{~b}\nabla^ca(a^{-1}\lambda_{bc}\dot{)}.
\end{equation}
Using (\ref{omega}) and (\ref{comm}) we get the evolution of $r_a$,
\begin{equation}
\label{raR} r_a=R_aa^{-3+3w}~,~~~~ \dot{R}_a=0.
\end{equation}

The other three variables $\sigma_{ab}$, $E_{ab}$ and $H_{ab}$ are tracefree symmetric tensors. Equations (\ref{lsdot})-(\ref{lhdot}) can be recast in the form,

\begin{widetext}
\begin{eqnarray}
& & \label{tris} \triangle \sigma_{ab} +\frac{5}{3}\theta\dot{\sigma}_{ab}+\frac{1-3w}{6}\theta^2\sigma_{ab} = \nabla_{\langle a}(a^{-2}(a^2\du_{b\rangle}\dot{)})-\nabla_{\langle a}Z_{b\rangle} -2\nabla_{\langle a}r_{b\rangle}, \\
& & \label{trie} \triangle E_{ab} +\frac{7}{3}E_{ab}+\fc(1-w)\theta^2E_{ab}-\frac{(1+w)(1+3w)}{18}\theta^2\sigma_{ab} = -\fa(1-w)\nabla_{\langle a}X_{b\rangle}, \\
& & \label{trih} \triangle H_{ab} +\frac{7}{3}H_{ab}+\fc(1-w)\theta^2H_{ab} = -\frac{1}{6}(1+w)\theta^2\nabla_{\langle a}\omega_{b\rangle},
\end{eqnarray}
where, $\triangle\lambda_{ab}=\ddot{\lambda}_{ab}-\lp\lambda_{ab}=\ddot{\lambda}_{ab}-h_{a}^{~c}h_{b}^{~d}h^{pq}\nabla_p (h_c^{~r}h_d^{~s}h_q^{~t}\nabla_t \lambda_{rs})$. In deriving (\ref{tris})-(\ref{trih}), we have used following identities for $B_{ab}=h_{(a}^{~c}h_{b)}^{~d}B_{cd}$ and $v_a=h_{a}^{~b} v_b$:
\begin{equation}
 (\mc B_{ab}\dot{)}=\mc a(a^{-1}B_{ab}\dot{)},\quad (\nabla_{\langle a}v_{b\rangle}\dot{)}=\nabla_{\langle a}a(a^{-1}v_{b\rangle}\dot{)}, 
\end{equation}
\begin{eqnarray}
 \mc^2 B_{ab}=\fa(h_{ab}\lp-h_{(a}^{~c}h_{b)}^{~d}\nabla_c\nabla_d)B^e_{~e}+\frac{3}{2}\nabla_{\langle a}h_{b\rangle c}\nabla^dB_d^{~c}+2h_{(a}^{~c}h_{b)}^{~d}\nabla_d\nabla^eB_{[ce]} - \lp B_{(ab)}.
\end{eqnarray}
\end{widetext}

The homogeneous parts of the differential equations (\ref{tris})-(\ref{trih}), obtained by setting $X_a, Z_a, \omega_a$ equal to zero give the pure tensor perturbations or gravitational waves \cite{Hawking}, \cite{Dunsby:1998hd}. However there are inhomogeneous parts of those equations. We have presented the solution of the inhomogeneous equation (\ref{tris}) for $\sigma_{ab}$, using $X_a$, $Z_a$ and $\omega_a$ as source. In particular, we will consider superhorizon modes which are responsible for large scale inhomogeneities.


\subsection{Radiation,  ``$w=\fb$"}

For radiation dominated collapsing FLRW background [$a=(-t/t_*)^{1/2}$, $\theta=3/2t$], (\ref{ddtX}) takes the form,
\begin{equation}
\label{XaR} \ddot{X}_a+\frac{11}{2t}\dot{X}_a+\frac{9}{2t^2}X_a-\fb h_a^{~b}\nabla_b\nabla^c X_c=0.
\end{equation}
Since $X_a$ is constructed from a scalar $\mu$, we can expand $X_a$ in Fourier modes in terms of spatial harmonics, defined in the Appendix,
\begin{equation}
 X_a = \sum_k X(k,t)Q^{(0)}_{a}.
\end{equation}
The last term in the left-hand side of (\ref{XaR}) reduces to,
\begin{eqnarray}
 \fb h_a^{~b}\nabla_b\nabla^c X_c &=& -\fb\sum_k X(k,t) \frac{a}{k} h_a^{~b}\nabla_b \nabla^ch_c^{~d}\nabla_d Q^{(0)} \nonumber \\
          &=& -\fb\sum_k\frac{k^2}{a^2}X(k,t)Q^{(0)}_a.
\end{eqnarray}
Thus we obtain an equation for Fourier mode $X(k,t)$,
\begin{equation}
\label{XR} \ddot{X}(k,t)+\frac{11}{2t}\dot{X}(k,t)+\left(\frac{9}{2t^2}+\frac{k^2}{3a^2}\right)X(k,t)=0.
\end{equation}
The first term in the coefficient of $X$ in (\ref{XR}) grows as $t^{-2} \sim a^{-4}$, whereas the second term grows as $a^{-2}$. So except for some very short wavelength modes the first term dominates over the second, near the bounce.

In terms of conformal time,
\begin{equation}
\label{conf} \eta = -\int \frac{dt}{a}=-2t_*a,
\end{equation}
the general solution of (\ref{XR}) becomes
\begin{equation}
 X(k,\eta)=\eta^{-\frac{9}{2}}\left( C_1(k) Y_{\frac{3}{2}}\left(-\frac{k\eta}{\sqrt{3}}\right)+C_2(k) J_{\frac{3}{2}}\left(-\frac{k\eta}{\sqrt{3}}\right)\right),
\end{equation}
where $J$ is Bessel function and $Y$ is Neumann function.

For superhorizon modes,
\begin{equation}
 \frac{k}{a}\ll\left|H\right| \Rightarrow 2kt_*a\ll 1.
\end{equation}
In that limit, $-\frac{k\eta}{\sqrt{3}}=\frac{2kt_*a}{\sqrt{3}}<<1$, we can use the asymptotic expansion of Bessel functions for a small argument.
\begin{equation}
\label{Xktr} X(k,t)=X^{(1)}(k)a^{-6}+X^{(2)}(k)a^{-3}.
\end{equation}
 Then, using (\ref{lxdot}), we obtain
\begin{equation}
 \label{Zktr} Z(k,t)=Z^{(1)}(k)a^{-4}+Z^{(2)}(k)a^{-1},
\end{equation}
where,
\begin{equation}
 Z^{(1)}(k)= \fc\theta_*^{-1}X^{(1)}(k),~~ Z^{(2)}(k)= -\fb\theta_*^{-1}X^{(2)}(k).
\end{equation}
$\theta_*$ is the value of expansion at initial time surface $t=-t_*$.

The equation (\ref{tris}) for radiation dominated background takes the form, 
\begin{widetext}
\begin{eqnarray}
\label{ddts} \ddot{\sigma}_{ab} + \frac{5}{3}\theta\dot{\sigma}_{ab}-\lp\sigma_{ab} = -\frac{1}{4\kappa M}\nabla_{\langle a}(a^{-2}(a^6X_{b\rangle}\dot{)})-\nabla_{\langle a}Z_{b\rangle}-2\nabla_{\langle a}r_{b\rangle}.~
\end{eqnarray}
\end{widetext}
$\sigma_{ab}$ is a traceless symmetric tensor and one can decompose it in three parts,
\begin{eqnarray}
\label{sTVS} \sigma_{ab} &=& \sigma^T_{ab} + \sigma^V_{ab} + \sigma^S_{ab} \nonumber \\
 &=& \sum_k \sigma^T Q^{(2)}_{ab} + \sum_k \sigma^V Q^{(1)}_{ab} + \sum_k \sigma^S Q^{(0)}_{ab}.
\end{eqnarray}
First two terms in the right-hand side of Eq.~(\ref{ddts}) are constructed from scalars, whereas the third term is originated from a divergenceless vector.
\begin{eqnarray}
\nabla_{\langle a}(a^{-2}(a^6X_{b\rangle}\dot{)}) &=& \sum_k (a^{-2}(a^6X(k,t)\dot{)}) \nabla_{\langle a}Q^{(0)}_{b \rangle} \nonumber \\ 
\label{XS} &=& -\sum_k k\theta_* X^{(2)}(k)a^{-2}Q^{(0)}_{ab}. 
\end{eqnarray}
\begin{eqnarray}
 \nabla_{\langle a}Z_{b\rangle} &=& \sum_k Z(k,t) \nabla_{\langle a}Q^{(0)}_{b \rangle} \nonumber \\
\label{ZS} = &-& \sum_k k\left( Z^{(1)}(k)a^{-5}+Z^{(2)}(k)a^{-2} \right)Q^{(0)}_{ab}.
\end{eqnarray}
From (\ref{raR}) we can expand $r_a$ in Fourier modes for $w=\fb$,
\begin{equation}
 r_a = \sum_k r(k,t) Q^{(1)}_a = \sum_k R(k)a^{-2} Q^{(1)}_a,
\end{equation}
\begin{eqnarray}
\label{rV} \nabla_{\langle a}r_{b\rangle} &=& \sum_k r(k,t) \nabla_{\langle a}Q^{(1)}_{b \rangle} 
          = -\sum_k kR(k)a^{-3}Q^{(1)}_{ab}.~~
\end{eqnarray}
Using the Fourier expansion, (\ref{sTVS}),(\ref{XS}), (\ref{ZS}) and (\ref{rV}), the Eq.~(\ref{ddts}) can be decomposed into three parts,
\begin{eqnarray}
\label{sT} & & \ddot{\sigma}^T(k,t)+\frac{5}{3}\theta\dot{\sigma}^T(k,t)+\frac{k^2}{a^2}\sigma^T(k,t) = 0, \\
\label{sV} & & \ddot{\sigma}^V(k,t)+\frac{5}{3}\theta\dot{\sigma}^V(k,t)+\frac{k^2}{a^2}\sigma^V(k,t) = V(k)a^{-3}, \\
& & \ddot{\sigma}^S(k,t)+\frac{5}{3}\theta\dot{\sigma}^S(k,t)+\frac{k^2}{a^2}\sigma^S(k,t)   \nonumber \\
\label{sS} & & \quad \quad \quad \quad \quad \quad \quad \quad \quad = S^{(1)}(k)a^{-5} + S^{(2)}(k)a^{-2},
\end{eqnarray}
where
\begin{equation}
 V=2kR,~S^{(1)}= \frac{2k}{3\theta_*}X^{(1)},~S^{(2)}=\frac{5k}{12\theta_*}X^{(2)}.
\end{equation}
Using conformal time (\ref{conf}), the propagation of gravitational waves are obtained from (\ref{sT}) \cite{Dunsby:1998hd},
\begin{equation}
 \sigma^T(k,\eta)= \eta^{-\frac{3}{2}}\left( D_1(k) Y_{\frac{3}{2}}\left(-k\eta\right)+D_2(k) J_{\frac{3}{2}}\left(-k\eta\right)\right).
\end{equation}
In the superhorizon limit $-k\eta=2kt_*a\ll 1$,
\begin{equation}
 \sigma^{T}(k,t) = \Sigma^{(1)}_T(k)a^{-3}+\Sigma^{(2)}_T(k).
\end{equation}
In this limit we can omit the $\frac{k^2}{a^2}$ term in (\ref{sT})-(\ref{sS}). Then the solutions of (\ref{sV}) and (\ref{sS}) are
\begin{eqnarray}
 \sigma^{V}(k,t) &=& \Sigma_V(k)a, \\
 \sigma^{S}(k,t) &=& \Sigma^{(1)}_S(k)a^{-1}+\Sigma^{(2)}_S(k)a^2.
\end{eqnarray}
Propagation of acceleration $\du_a$ and its spatial derivatives $A$, $A_a$ are obtained from (\ref{uadot}),
(\ref{AA}) and (\ref{Xktr}),
\begin{eqnarray}
 \du_a &=& -\frac{3}{4}\theta_*^{-2}\sum_k \left( X^{(1)}(k)a^{-2}+X^{(2)}(k)a^{} \right)Q^{(0)}_a, \\
 A &=& -\frac{3}{4}\theta_*^{-2}\sum_k k\left( X^{(1)}(k)a^{-3}+X^{(2)}(k) \right)Q^{(0)}, \\
 A_a &=& \frac{3}{4}\theta_*^{-2}\sum_k k^2\left( X^{(1)}(k)a^{-4}+X^{(2)}(k)a^{-1} \right)Q^{(0)}_a.~~~ 
\end{eqnarray}


\subsection{Dust, ``$w=0$''}

In a dust dominated background, the density perturbation is scale invariant, as evident from (\ref{ddtX}). Also the acceleration is zero. The behavior of $X_a$ and $Z_a$ are \cite{Ellis:1989jt}
\begin{eqnarray}
\label{Xktd} X(k,t) &=& X^{(1)}(k)a^{-11/2}+X^{(2)}(k)a^{-3},\\
\label{Zktd} Z(k,t) &=& Z^{(1)}(k)a^{-4}+Z^{(2)}(k)a^{-3/2}.
\end{eqnarray}
However, the shear is not scale invariant. For superhorizon wavelengths, pure tensor components of $\sigma_{ab}$ are \cite{Dunsby:1998hd}
\begin{equation}
 \sigma^{T}(k,t) = \Sigma^{(1)}_T(k)a^{-3}+\Sigma^{(2)}_T(k)a^{-1/2}.
\end{equation}
The vector and scalar components of $\sigma_{ab}$ are found to be
\begin{eqnarray}
 \sigma^{V}(k,t) &=& \Sigma_V(k)a^{-1}, \\
 \sigma^{S}(k,t) &=& \Sigma^{(1)}_S(k)a^{-3}\log a+\Sigma^{(2)}_S(k)a^{1/2}.
\end{eqnarray}


\section{Comparison of linear and nonlinear terms}
\label{compare}

We have seen that each variable has a growing mode(s) near the bounce. But this growing mode(s) can be absorbed by a mere redefinition of that variable. For example, if a variable $f$ behaves as $f\sim a^{-n}$ near the bounce, then we can construct a new variable $\tilde{f}=a^mf$, such that $m\ge n$, which remains finite. So the growing modes of perturbations do not rule out their linear evolution. In order to investigate whether the perturbations remain linear near the bounce one needs to compare linear and nonlinear terms of the perturbation equations using the behavior of linearity parameters defined in Sec.~\ref{condlin}.


\subsection{Radiation dominated case}

Near the bounce, we will consider only the dominating modes of different variables. 
\begin{eqnarray}
& & X_a = \bar{X}_a a^{-6}, \;\;\; Z_a=\bar{Z}_a a^{-4}, \\
& & \du_a = U_a a^{-2}, \;\;\; A=\bar{A}a^{-3}, \;\;\; A_a=\bar{A}_a a^{-4}, \\
& & \omega_{ab}=\Omega_{ab}a^{-1}, \;\;\; \sigma_{ab}=\Sigma_{ab}a^{-3}.
\end{eqnarray}
The higher order value of curvature perturbation (\ref{r}) is
\begin{eqnarray}
 \mathcal{R} &=& A+2(\omega^2-\sigma^2) \nonumber \\
 &=& \bar{A}a^{-3} +2\Omega^2a^{-2}-2\Sigma^{2}a^{-6} \nonumber \\
&=& -\bar{\mathcal{R}}a^{-6},
\end{eqnarray}
where, $\bar{X}_a$, $\bar{Z}_a$, $U_a$, $\bar{A}$, $\bar{A}_a$, $\Omega_{ab}$, $\Sigma_{ab}$ and $\bar{\mathcal{R}}$ are time independent.

Using the commutation of time and spatial derivatives,
\begin{eqnarray}
 h_a^{~b}\nabla_b\omega^2=\Omega_a a^{-3},~~~h_a^{~b}\nabla_b\sigma^2=\Sigma_a a^{-7},
\end{eqnarray}
with $\dot{\Omega}_a=\dot{\Sigma}_a=0$. Now we can calculate the linearity parameters defined in (\ref{eps}),
\begin{eqnarray}
& & \varepsilon_1 = \frac{\left|\Omega^b_{~a}\bar{X}_b\right|}{\left|\frac{4}{3}\kappa M\bar{Z}_a\right|}a,~~~
    \varepsilon_2 = \frac{\left|\Sigma^b_{~a}\bar{X}_b\right|}{\left|\frac{4}{3}\kappa M\bar{Z}_a\right|}a^{-1},~~~\\
& & \varepsilon_3 = \frac{\left|\bar{\mathcal{R}}U_a\right|}{\left|\fa \bar{X}_a\right|}a^{-2},~
    \varepsilon_4 = \frac{\left|2\Omega_a\right|}{\left|\fa \bar{X}_a\right|}a^3,~
    \varepsilon_5 = \frac{\left|2\Sigma_a\right|}{\left|\fa \bar{X}_a\right|}a^{-1},~~~~\\
& & \varepsilon_6 = \frac{\left|\Omega^b_{~a}\bar{Z}_b\right|}{\left|\fa \bar{X}_a\right|}a,~
    \varepsilon_7 = \frac{\left|\Omega^b_{~a}\bar{Z}_b\right|}{\left|\fa \bar{X}_a\right|}a^{-1},~ \\
& & \tilde{\varepsilon}_3 = \frac{\left|\bar{\mathcal{R}}U_a\right|}{\left|\bar{A}_a\right|}a^{-4},~
    \tilde{\varepsilon}_4 = \frac{\left|2\Omega_a\right|}{\left|\bar{A}_a\right|}a,~~
    \tilde{\varepsilon}_5 = \frac{\left|2\Sigma_a\right|}{\left|\bar{A}_a\right|}a^{-3},~~~\\
& & \tilde{\varepsilon}_6 = \frac{\left|\Omega^b_{~a}\bar{Z}_b\right|}{\left|\bar{A}_a\right|}a^{-1},~
    \tilde{\varepsilon}_7 = \frac{\left|\Omega^b_{~a}\bar{Z}_b\right|}{\left|\bar{A}_a\right|}a^{-3}.~ 
\end{eqnarray}

Let at some time slice $t=-t_1$, such that $t_*\gg t_1\gg t_b$, linearity conditions are satisfied. So $\varepsilon_2(-t_1)=\frac{\left|\Sigma^b_{~a}\bar{X}_b\right|}{\left|\frac{4}{3}\kappa M\bar{Z}_a\right|}a_1^{-1} \ll 1$. If we consider another time slice $t=-t_2$, which is close to $t_b$, i. e. $t_1\gg t_2\gtrsim t_b$, then
\begin{equation}
 \varepsilon_2(-t_2)=\varepsilon_2(-t_1)\frac{a_1}{a_2}.
\end{equation}
Since $a_2\ll a_1$, the parameter $\varepsilon_2$ may become order $1$ at $t=-t_2$ and the condition (1) no longer holds. Similar arguments can be given for $\varepsilon_3$, $\varepsilon_5$, $\varepsilon_7$, $\tilde{\varepsilon}_3$, $\tilde{\varepsilon}_5$, $\tilde{\varepsilon_6}$ and $\tilde{\varepsilon_7}$.


\subsection{Dust dominated case}

The evolution of variables in dust dominated collapsing FLRW background near the bounce,
\begin{eqnarray}
 & & X_a = \bar{X}_a a^{-11/2}, \;\;\; Z_a=\bar{Z}_a a^{-4,}. \\
& & \omega_{ab}=\Omega_{ab}a^{-2}, \;\;\; \sigma_{ab}=\Sigma_{ab}a^{-3}\log a, \\
& & h_a^{~b}\nabla_b\omega^2=\Omega_a a^{-5},\;\;\; h_a^{~b}\nabla_b\sigma^2=\Sigma_a a^{-7}|\log a|^2.
\end{eqnarray}
Since $A_a=0$, $\tilde{\varepsilon}_3$-$\tilde{\varepsilon}_7$ are undefined. So, to preserve linearity, all $\varepsilon_1$-$\varepsilon_7$ must be much less than $1$. 
\begin{eqnarray}
& & \varepsilon_1 = \frac{\left|\Omega^b_{~a}\bar{X}_b\right|}{\left|\kappa M\bar{Z}_a\right|}a^{-\fa},~~~
    \varepsilon_2=\frac{\left|\Sigma^b_{~a}\bar{X}_b\right|}{\left|\kappa M\bar{Z}_a\right|}a^{-\frac{3}{2}}|\log a|~~~\\
& & \varepsilon_3 = 0,~
    \varepsilon_4 = \frac{\left|2\Omega_a\right|}{\left|\fa \bar{X}_a\right|}a^{\fa},~
    \varepsilon_5 = \frac{\left|2\Sigma_a\right|}{\left|\fa \bar{X}_a\right|}a^{-\frac{3}{2}}|\log a|^2~~~~\\
& & \varepsilon_6 = \frac{\left|\Omega^b_{~a}\bar{Z}_b\right|}{\left|\fa \bar{X}_a\right|}a^{-\fa},~
    \varepsilon_7 = \frac{\left|\Omega^b_{~a}\bar{Z}_b\right|}{\left|\fa \bar{X}_a\right|}a^{-\frac{3}{2}}|\log a|.
\end{eqnarray}
So, in this case also some of the parameters $\varepsilon_1$, $\varepsilon_2$, $\varepsilon_5$, $\varepsilon_6$, $\varepsilon_7$ may become order $1$, near the bounce.


\section{Conclusion} 

Previously it has been shown \cite{Allen:2004vz}, \cite{Vitenti:2011yc} that in some commonly defined gauges linear perturbations grow so much near the bounce that that may invalidate the linear perturbation theory. But the perturbations remain small if one uses some other gauge fixing condition. It has been argued that the commonly defined gauges are not well defined near and at the bounce. Hence linear perturbation theory is valid if one uses only the well-defined gauge. Naturally, a question arises whether these results are real or gauge artifacts.

In order to investigate this issue we used the covariant approach to perturbation theory. We focus on the evolution equations for density perturbation $X_a$. The validity conditions of linear approximation of the (nonlinear) density perturbation equations are set in terms of some linearity parameters. Then the linear perturbation equations are solved for a collapsing FLRW background near the bounce. The solutions are used to compute the linearity parameters. It is found that some of those parameters grow beyond order unity near the bounce in both radiation and dust dominated cases. That means the nonlinear terms are comparable to the linear terms. So unless some special initial conditions are imposed on the variables such as shear and vorticity, perturbations may not be linear near the bounce.

Thus we conclude that perturbations may not be linear near the bounce and linear perturbation theory may not be adequate to give proper evolution of perturbations through the bounce. In sharp contrast with the result obtained in \cite{Vitenti:2011yc} our result is independent of choice of gauge. We used gauge invariant variables that were not assumed to be small with respect to background. So one can evolve them through the bounce and match with corresponding quantities in the expanding phase---but this would require the full nonlinear analysis.

In this work, we consider only the contracting branch and used general relativity with usual matter distribution as a correct theory to describe the dynamics of the universe. To investigate the nonlinearity of perturbations in a concrete manner, we have to take specific models of bounce. Currently we are working on this issue and hope to report in future.

\acknowledgments
I would like to thank Amit Ghosh for encouragement and guidance. Financial assistance is given by the Council of Scientific and Industrial Research, Government of India.


\appendix
\section{Spatial Harmonics}
\label{harmonics}

The tensor eigenfunctions (harmonics) of the spatial Laplacian $\lp=h^{ab}\nabla_ah_b^{~c}\nabla_c$ listed below\cite{Bardeen:1980kt},\cite{Goode:1989jt}, are solutions of the tensor Helmholtz equation.
\begin{equation}
 \lp Q_{ab...c} + \frac{k^2}{a^2}Q_{ab...c}=0.
\end{equation}

(1) \textbf{Scalar harmonics}: Harmonics constructed from solutions of the scalar Helmholtz equation, 
\begin{equation}
 \lp Q^{(0)} + \frac{k^2}{a^2}Q^{(0)}=0.
\end{equation}
Vector and tensor eigenfunctions constructed from the scalars are
\begin{equation}
 Q^{(0)}_a= -\frac{a}{k}h_a^{~b}\nabla_bQ^{(0)},
\end{equation}
\begin{eqnarray}
 Q^{(0)}_{ab} &=& -\left(\frac{a}{k}\right) \nabla_{\langle a}Q^{(0)}_{b \rangle}\nonumber \\
              &=& \left(\frac{a}{k}\right)^2 \nabla_{\langle a}\nabla_{b \rangle}Q^{(0)}\nonumber \\
              &=& \left(\frac{a}{k}\right)^2 h_{(a}^{~c}h_{b)}^{~d}\nabla_c\nabla_dQ^{(0)}+\fb h_{ab}Q^{(0)}.
\end{eqnarray}

(2) \textbf{Vector Harmonics}: Harmonics constructed from solutions of the vector Helmholtz equation, 
\begin{equation}
 \lp Q^{(1)}_a + \frac{k^2}{a^2}Q^{(1)}_a=0, ~~~~ \nabla^a Q^{(1)}_a=0.
\end{equation}
Tensor eigenfunctions constructed from the vectors are

\begin{eqnarray}
 Q^{(1)}_{ab} = -\left(\frac{a}{k}\right) \nabla_{\langle a}Q^{(1)}_{b \rangle}.
\end{eqnarray}

(3) \textbf{Tensor Harmonics}: Harmonics constructed from solutions of the tensor Helmholtz equation, 
\begin{equation}
 \lp Q^{(2)}_{ab} + \frac{k^2}{a^2}Q^{(2)}_{ab}=0, ~~ \nabla^b Q^{(2)}_{ab}=0, ~~ Q^{(2)a}_{~a}=0.
\end{equation}


\end{document}